\documentclass[twocolumn,showpacs,superscriptaddress,amsmath,amssymb,prl]{revtex4-1}
\usepackage[normalem]{ulem}
\usepackage{epsfig}
\usepackage{amsfonts}
\usepackage{amsmath}
\usepackage{slashed}
\usepackage{graphicx}
\usepackage{color}
\usepackage{mathtools}
\usepackage{subfigure}
\usepackage{graphicx}

\usepackage{stmaryrd}
\usepackage{amssymb,psfrag}
\usepackage[normalem]{ulem}
\usepackage{caption}

\newcommand{\beq}{\begin{eqnarray}}
\newcommand{\eeq}{\end{eqnarray}}

\newcommand{\nn}{\nonumber}

\begin{document}
\title{Probing the Core of Nuclear Structure through the $\pi N$ Scattering at an Electron-Positron Collider}

\author{Wei Wang}
\email{wei.wang@sjtu.edu.cn}
\affiliation{State Key Laboratory of Dark Matter Physics, Key Laboratory for Particle Astrophysics and Cosmology (MOE), Shanghai Key Laboratory for Particle Physics and Cosmology,
School of Physics and Astronomy, Shanghai Jiao Tong University, Shanghai 200240, China}

\author{Ji Xu}
\email{xuji@lzu.edu.cn}
\affiliation{School of Nuclear Science and Technology, Lanzhou University, Lanzhou 730000, China}

\author{Ya-Teng Zhang}
\email{zhangyateng@zzu.edu.cn}
\affiliation{School of Physics and Microelectronics, Zhengzhou University, Zhengzhou, Henan 450001, China}

\author{Xiao-Rong Zhou}
\email{zxrong@ustc.edu.cn}
\affiliation{University of Science and Technology of China, Hefei 230026, China}
\affiliation{State Key Laboratory of Particle Detection and Electronics, Hefei, 230026, China}

\date{\today}

\begin{abstract}
Short-range correlation pairs (SRCs)---core of nuclear structure, composed of highly off-shell nucleons---are mostly studied via electron-nucleon scattering, leaving a gap in meson-based probes. We propose probing SRC off-shell nucleons via quasielastic $\pi^+$-bound proton scattering ($\pi^+ p \to \pi^+ p$) at electron-positron colliders, of which the beryllium-based ($^{9}$Be) beam pipe of the BESIII experiment operating at BEPCII, addresses a key gap and enables meson-beam investigations of SRCs. We point out that off-shellness of SRC nucleons yields measurable signatures: accumulated missing energy ($\sim0.1$\,GeV), shifted proton effective mass (0.7-0.8\,GeV), and cross-section differences from free scattering or with only Fermi motion. As an estimate, we find that BESIII's high luminosity and $\pi^+$ yield support $\sim10^4$ scattering events, while  STCF ($50\times$ higher luminosity) will greatly  enhance this number. This first meson-beam SRC study at an electron-positron collider fills $^{9}$Be research gaps and advances understanding of nuclear structure core and nonperturbative QCD.
\end{abstract}

\maketitle

\textit{Introduction.}---The structural integrity of atomic nuclei is upheld by the bound protons ($p$) and neutrons ($n$), collectively referred to as nucleons ($N$). For over half a century, the nuclear shell model has served as the standard and dominant framework for describing atomic nuclei. In this model, nucleons move independently in well-defined quantum orbits under the influence of an average mean field generated by their mutual interactions. While mean field calculations have proven to be highly useful, they do not capture the complete structure of nuclei, particularly the effects of small, dense structures known as short-range correlation pairs (SRCs).

In atomic nuclei, nucleons can occasionally undergo strong, short-range interactions with each other, ending up in a configuration with large relative momenta but small total momentum. These naturally transient high-density fluctuations are SRCs \cite{Frankfurt:1988nt,Arrington:2011xs,Hen:2016kwk,Miller:2020eyc}.  {SRCs are a core component in the unified description of nuclear structure---one of the central objectives of modern nuclear physics. These structures consist mainly of highly off-shell nucleons, characterized by an energy-momentum relation that significantly deviates from that of free protons and neutrons.} Consequently, since their initial discovery \cite{Frankfurt:1993sp}, a continuous stream of related theoretical and experimental studies emerged \cite{Tang:2002ww,Hen:2014nza,Subedi:2008zz,Fomin:2011ng,CLAS:2020mom,CLAS:2005ola,Li:2022fhh,Zhang:2025nst,Xu:2019wso,Hatta:2019ocp,Ma:2023tsi,Sargsian:2012sm,Wang:2024ikx}. However, a theoretical consensus on SRCs has yet to be established. The nucleon momentum distributions exhibit a high-momentum tail which is similar for all nuclei from deuteron to heavy nulcei. This universality provides strong evidence against the role of the collective or mean field effects. The interaction between nucleons arises from the residual strong force within QCD, which cannot be described by perturbative QCD methods. Furthermore, nonperturbative methods such as lattice QCD also remain impractical on this issue, as nuclear structure constitutes a prototypical many-body problem that currently exceeds the computational capabilities for lattice simulations \cite{Detmold:2020snb,Chen:2024rgi}. In addition, deep inelastic scattering (DIS) experiments have revealed that the quark distributions in nucleons bound within nuclei differ from those in free nucleons---a phenomenon known as the EMC effect \cite{EuropeanMuon:1983wih,EuropeanMuon:1988lbf}. This EMC effect were recently associated with the large off-shellness nucleons in SRC pairs \cite{Weinstein:2010rt,CLAS:2019vsb,CLAS:2018xvc}, which has garnered increasing attentions \cite{Chen:2016bde,Wang:2020uhj,Arrington:2012ax,Malace:2014uea,Wang:2024cpx}, as SRCs are regarded as a key transitional structure connecting the quark-gluon degrees of freedom with the nucleonic degrees of freedom \cite{nCTEQ:2023cpo}. The currently available experimental data are insufficient; therefore, further experiments are required to investigate the characteristics of SRC.

At present, experimental knowledge of SRCs is derived almost exclusively from proton and electron-nucleon scattering at SLAC, Brookhaven and Jefferson Laboratories \cite{Frankfurt:1993sp,Tang:2002ww,Hen:2014nza,Subedi:2008zz,Fomin:2011ng,CLAS:2020mom,CLAS:2005ola,Li:2022fhh,Zhang:2025nst}. To advance this research, the utilization of a broader range of probes, such as incident meson beams ($\pi$, $K$, etc.), is warranted. Given that the cross section of pion-nucleon scattering is typically two to three orders of magnitude larger than that of electron-nucleon scattering, making the pion as an ideal probe for studding SRC. As one of the most fundamental processes in hadron and nuclear physics, $\pi N$ scattering has been intensively studied with a long history in theory and experiment \cite{Bernard:1995dp,Ellis:1997kc,Bernard:2007zu}. The pion, being the lightest strongly interacting hadron, constitutes an essential component in all theories of nucleon-nucleon interaction. Previous studies on $\pi N$ scattering have primarily focused on the fundamental coupling constants in chiral perturbation theory (ChPT) and the structure of resonant states, no experimental study has utilized pion beams to probe SRCs in target nuclei. Nevertheless, this method is highly viable and can serve as an important method in future research. Furthermore, experimental research on SRCs has so far been primarily conducted by utilizing $^{3}$H, $^{3}$He, $^{4}$He, $^{12}$C, $^{56}$Fe, $^{197}$Au, $^{208}$Pb, etc., with only few experiment employing $^{9}$Be \cite{Fomin:2011ng}. However, it is noteworthy that experimental data for beryllium---a kind of important nuclei whose density is similar to helium, yet whose EMC effect closely resembles that of denser nuclei due to its significant cluster structure---remains particularly valuable.

In this Letter, we aim to use pions to take a ``snapshot'' of nuclear configurations. We explore the potential of $\pi^+$-bound proton scattering for probing SRCs in an electron-positron collision experiment such as the Beijing Spectrometer III (BESIII) \cite{BESIII:2009fln}. The continuous pion beam with feasible energies, along with the high luminosity and high-performance detectors available at BESIII, is expected to be leveraged to its full potential. Thanks to the beryllium ($^{9}$Be) beam pipe placed closely adjacent to the $e^+e^-$ beams, we can utilize the protons within it as the target. Since the proposal in \cite{Yuan:2021yks}, studies utilizing the beam pipe at BESIII have come under the limelight \cite{BESIII:2023clq,BESIII:2023trh,BESIII:2024geh}. Due to the existence of intermediate resonant states (such as $\Delta(1232)$ and $N^*$), which lie within the energy range covered by BESIII, the cross section of $\pi^+$-bound proton scattering is significantly enhanced. Therefore, we have good reason to expect that a substantial dataset can be acquired and systematic uncertainties effectively controlled. Consequently, BESIII can pioneer the first attempt to study SRC using a meson beam, and remarkably, this will be achieved at an electron-positron collider. We propose that the off-shellness of nucleons within SRCs will manifest as measurable differences in the \textit{missing energy spectrum, proton effective mass, and quasielastic cross section between $\pi^+$-bound proton and $\pi^+$-free proton scattering.} Leveraging these signatures enables a quantitative investigation of this core aspect of nuclear structure.

\begin{figure}[htbp]
  \includegraphics[width=1\columnwidth]{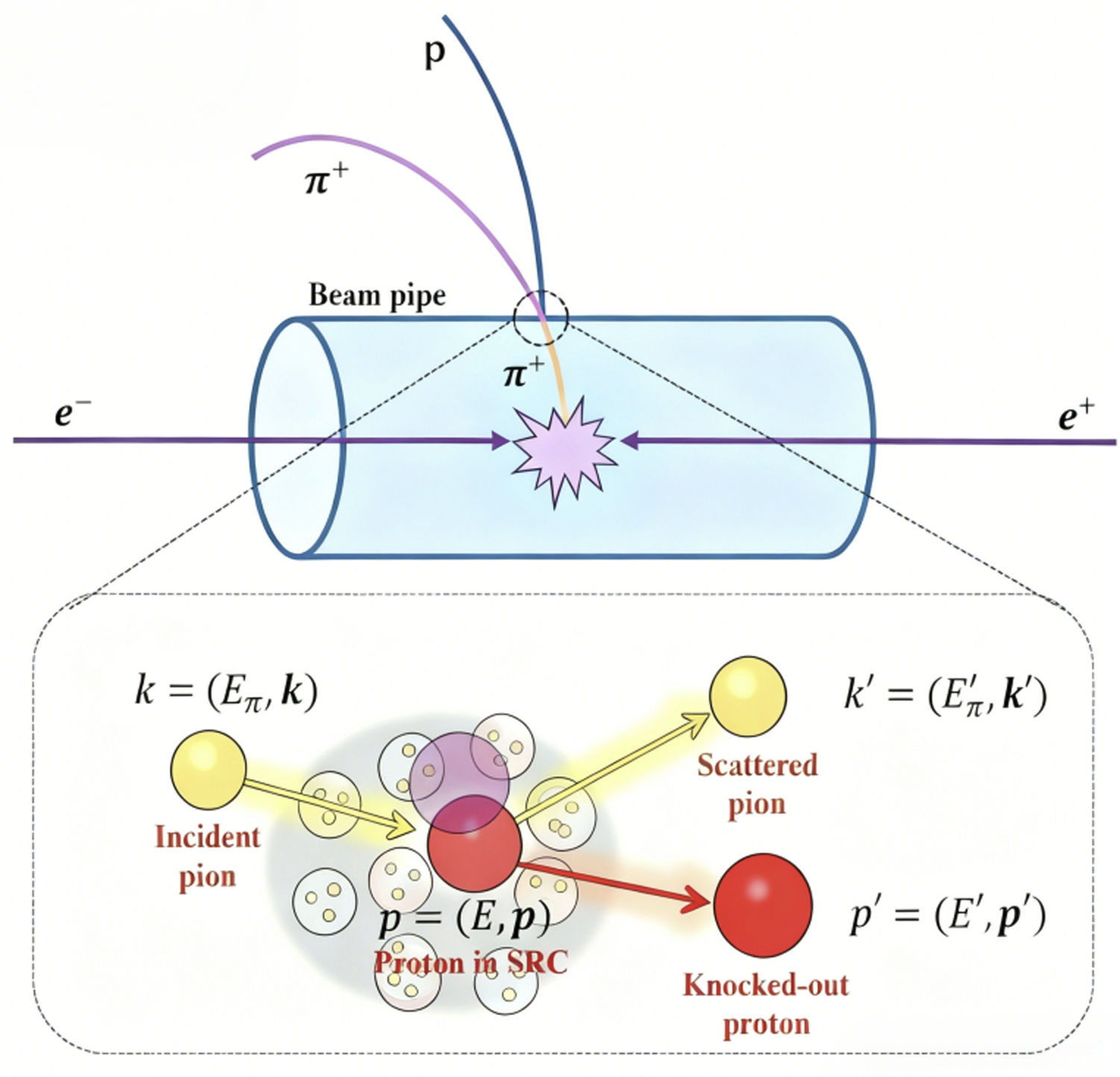}
  \caption{A pion is produced in electron-positron collision and subsequently strikes the beam pipe. Below is a schematic diagram of the scattering between the pion and a proton within the atomic nucleus of the beam pipe.}
  \label{schematic_v3}
\end{figure}

\textit{Studies of SRCs through the missing energy and effective mass of $\pi^+$-bound proton scattering.}---The process of interest is illustrated in Fig.\,\ref{schematic_v3}: a pion produced in the $e^+e^-$ collision strikes a proton within the beam pipe, which knocks the proton out of beryllium, leaving the rest of the system nearly unaffected (quasielastic scattering $\pi^+ p \to \pi^+ p$).

Here we label the initial and final momenta of pion as
\begin{eqnarray}
  && k = (E_\pi,\, \textbf{k}) \,, \qquad  k' = (E'_\pi,\, \textbf{k}')  \,.\nn
\end{eqnarray}
The momentum and energy transfer of the pion are $\textbf{q} \equiv \textbf{k}-\textbf{k}'$ and $\omega \equiv E_\pi-E'_\pi$. The proton in beryllium absorbs this transfer, we label its initial and final momenta as
\begin{eqnarray}
  && p = (E,\, \textbf{p}) \,, \qquad p' = (E',\, \textbf{p}') \,,\nn
\end{eqnarray}
with $\textbf{p}'=\textbf{p}+\textbf{q}$ and energy $E'=E+\omega$. Here, the initial state of the proton is the subject of our interest. If the initial proton were free, its experimentally measurable momentum would lie   on the mass shell. In contrast, if the initial proton was bound within nucleus, its wave function will be modified by the surrounding medium \cite{CiofidegliAtti:1995qe}, and this modification should exhibit dependence on the momentum of the proton \cite{CLAS:2020mom}. In beryllium, the majority of protons (about 80\%) have momenta below the Fermi momentum $k_F$ (we take the Fermi momentum of beryllium $k_F=0.206\,\textrm{GeV}$, see App.\,A for more details), their behaviors can be consistently described by the mean field calculations, and can be approximately treated as on-shell. While the minority of protons (about 20\% \cite{Subedi:2008zz}) belong to SRCs with high momentum and high off-shellness.

Therefore, we can define an experimentally measurable quantity: the missing energy $E_{\textrm{miss}}\equiv\omega-(E'-m_N)=m_N-E$, which quantifies the difference between the initial proton's energy and its rest mass, thereby characterizing its off-shell nature. On one hand, nearly all non-SRC protons can be treated (or approximately treated) as on-shell with $E_{\textrm{miss}}\simeq 0 \,\textrm{GeV}$. On the other hand, according to the results of electron-nucleon scattering data, when the momentum of proton bound in SRCs goes to $|\textbf{p}|=0.45\,\textrm{GeV}$, its missing energy reaches 0.1\,GeV. Therefore, it can be anticipated that by applying a momentum cutoff (e.g., $0.4\,\textrm{GeV}<|\textbf{p}|<0.5\,\textrm{GeV})$ in $\pi^+ p \to \pi^+ p$ process, BESIII will also observe an accumulation of events at $E_{\textrm{miss}} \simeq 0.1\,\textrm{GeV}$. Additionally, a linear relationship between $|\textbf{p}|$ and $E_{\textrm{miss}}$ should be observable. We derive the functional relationship between the proton's missing energy and its momentum
\begin{eqnarray}\label{TextEmissfunction}
  E_{\textrm{miss}}(|\textbf{p}|)=\left\{\begin{array}{cc}
0 \,, & \quad  |\textbf{p}|<k_F \\ \\
0.446 \, |\textbf{p}| -0.098  \,, & \quad  |\textbf{p}|>k_F
\end{array}\,,\right.
\end{eqnarray}
which can be validated by the future result of BESIII and will be useful for the subsequent calculations of $\pi^+$-bound proton scattering cross section.

In addition, the SRCs can also be investigated by measuring the effective mass. The binding of nucleons disrupts the relationship between the nucleon's mass, energy and momentum $E^2 \neq m_N^2 +|\textbf{p}|^2$, thereby generating a virtuality of the nucleon \cite{CiofidegliAtti:2007ork,Kim:2024wne,Xing:2023uhj}. Through measurements of $\pi^+$-bound proton scattering, BESIII can provide the distribution of proton virtuality within beryllium. Equivalently, the deformation of protons within SRCs is relatively large, leading to a deviation of their mass from $m_N \simeq 0.938\,\textrm{GeV}$. Therefore, they can also be studied experimentally by measuring the proton's effective mass $m_{N,\textrm{eff}}^2 = (E'+E'_\pi-E_\pi)^2-(\textbf{p}'+\textbf{k}'-\textbf{k})^2$. It is reasonable to expect that a peak or a broad shoulder would appear in the region of $m_{N,\textrm{eff}} \approx 0.7-0.8\,\textrm{GeV}$ (this corresponds to the protons in SRCs).

\begin{figure}[htbp]
\includegraphics[width=0.95\columnwidth]{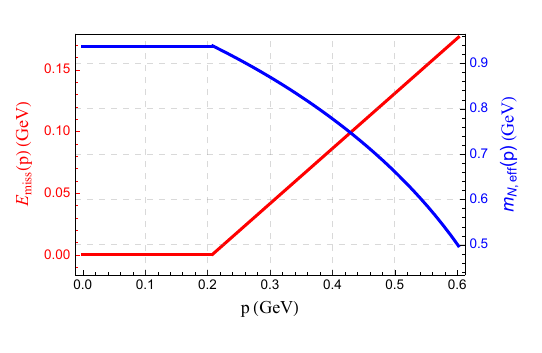}
\centering
\caption{The initial proton's missing energy $E_{\textrm{miss}}$ (red solid line) and effective mass $m_{N,\textrm{eff}}$ (blue solid line) as functions of its momentum. These two curves are represented by the legends corresponding to their respective colors.}
\label{EmissandMeff}
\end{figure}

Fig.\,\ref{EmissandMeff} presents the distributions of missing energy $E_{\textrm{miss}}$ and effective mass $m_{N,\textrm{eff}}$ as functions of the initial proton momentum. These observables are highly straightforward and experimentally accessible, and thus are recommended as priority measurements for future BESIII studies.

\textit{Studies of SRCs effects on the cross section of $\pi^+$-bound proton scattering.}---Free $\pi^+ p$ elastic scattering, being one of the most fundamental processes in particle and nuclear physics, has been extensively studied, leading to a precise determination of its cross section. When the proton is bound within a nucleus, this process provides a sensitive probe for studying SRCs. In this scenario, the finite momentum distribution and off-shellness of the bound proton imply that the $\pi^+$-bound proton scattering cross section is given by a convolution of the free cross section with the proton momentum distribution inside the nucleus,
\begin{eqnarray}\label{sigmabound1}
   \sigma_{\textrm{bound}}(|\textbf{k}|) = \int \frac{d^3 \textbf{p} }{(2\pi)^3} \, \sigma_{\textrm{free}}(\sqrt{s}) \, \rho_p(|\textbf{p}|) \,,
\end{eqnarray}
here $\sigma_{\textrm{free}}(\sqrt{s})$ is the free cross section with
\begin{eqnarray}
  \sqrt{s} &=& m_N^2 + m_\pi^2 + E_{\textrm{miss}}^2 - E_{\textrm{miss}} (2m_N + 2E_\pi) + 2m_N E_\pi \nn\\
  && - |\textbf{p}|^2 -2 |\textbf{k}| |\textbf{p}| \cos\theta \,.\nn
\end{eqnarray}
$\theta$ is the angle between the momenta of initial pion and proton.

Let us begin by discussing the free cross section. Extensive research involving free $\pi^+ p$ scattering (pions on free protons in $^1$H) and pion-deuteron scattering has been pursued, resulting in the accumulation of extensive datasets \cite{ParticleDataGroup:2024cfk}. Given this abundance of existing data, and considering the presence of numerous resonance states in $\pi^+ p$ scattering which makes a significant difficulty for theoretical predictions in a wide energy range ($1.08\,\textrm{GeV} \leq \sqrt{s} \leq 2.50\,\textrm{GeV}$), we choose to employ a data-fitting approach to derive the free cross section. See App.\,B for detailed expression of $\sigma_{\textrm{free}}(\sqrt{s})$.

We then move on to the proton momentum distribution $\rho_p(|\textbf{p}|)$, which incorporates contributions from both Fermi motion and SRCs. It is isotropic (do not depend on the direction of $\textbf{p}$) and obeys the normalization
\begin{eqnarray}
  Z = \int \frac{d^3 \textbf{p}}{(2\pi)^3} \, \rho_p(|\textbf{p}|) \,,\nn
\end{eqnarray}
where $Z=4$ is the number of protons in beryllium. It should be emphasized that this distribution can not be directly observed in experiments. We utilize the SRC dominance model to describe it \cite{Hen:2014nza,Sargsian:2012sm}:
\begin{eqnarray}\label{modelbySargsian}
  \rho_p(|\textbf{p}|)=\left\{\begin{array}{cc}
\eta \cdot \rho_p^{\textrm{MF}}(|\textbf{p}|) \,, & \quad  |\textbf{p}| < k_F \\\\
\frac{A}{2 Z} \cdot a_2(\textrm{Be}/d) \cdot \rho_d(|\textbf{p}|) \,, & \quad  |\textbf{p}| > k_F
\end{array}\,.\right.
\end{eqnarray}
Based on the magnitude of the proton momentum within beryllium, this model divides the distribution function into two parts. When $|\textbf{p}| < k_F$, this distribution is determined by the mean field nuclear wave-function calculations where $\rho_p^{\textrm{MF}}(|\textbf{p}|)$ is the mean field proton momentum distribution in beryllium. When $|\textbf{p}| > k_F$, this distribution is determined by the SRC scale factor $a_2(\textrm{Be}/d)$, which is an experimentally determined per-nucleon probability of finding a high-momentum nucleon in beryllium relative to deuteron. $\rho_d(|\textbf{p}|)$ is the deuteron momentum distribution, and $k_F=0.206\,\textrm{GeV}$ is the Fermi momentum of beryllium. $\eta$ serves as a normalization factor in order to guarantee $4\pi/(2\pi)^3\int_0^\infty \rho_{p}(|\textbf{p}|) |\textbf{p}|^2 d|\textbf{p}| =1$. The specific expression for the $\rho_p(|\textbf{p}|)$ in Eq.\,(\ref{modelbySargsian}) is provided in App.\,C.

\begin{figure}[htbp]
\includegraphics[width=1\columnwidth]{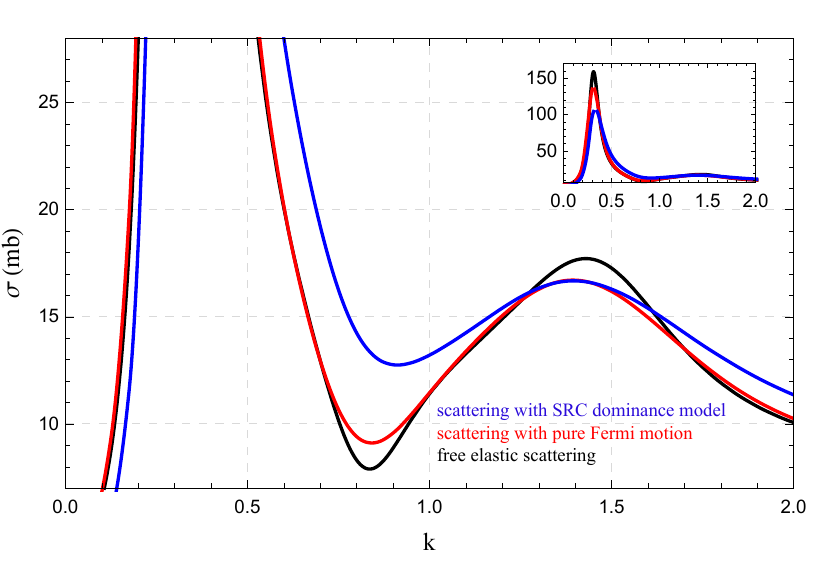}
\centering
\caption{Predictions on the cross section of $\pi^+ p$ scattering. The black line represents the experimental data for free $\pi^+ p$ elastic scattering. The blue line shows the result obtained by SRC dominance model. The red line shows the result considering only Fermi motion in Eq.\,(\ref{modelbySargsianpure}). The upper-right subplot depicts the behavior of the cross section at the $\Delta(1232)$ resonant state, illustrating that the nuclear correction reduces its magnitude.}
\label{boundsigma3}
\end{figure}

Substituting the expressions of $\rho_p(|\textbf{p}|)$ and $\sigma_{\textrm{free}}(\sqrt{s})$ into Eq.\,(\ref{sigmabound1}), we can obtain the cross section for $\pi^+$-bound proton scattering where the proton comes from beryllium. The blue line in Fig.\,\ref{boundsigma3} represents the $\sigma_{\textrm{bound}}(|\textbf{k}|)$ obtained by the SRC dominance model in Eq.\,(\ref{modelbySargsian}).

Additionally, we also want to examine how the bound cross section will be affected when only Fermi motion ($|\textbf{p}|<k_F$) is considered. We take the proton momentum distribution as:
\begin{eqnarray}\label{modelbySargsianpure}
  \rho_p^{\textrm{pure}}(|\textbf{p}|)=\left\{\begin{array}{cc}
\eta^{\textrm{pure}} \cdot \rho_p^{\textrm{MF}}(|\textbf{p}|) \,, & \quad  |\textbf{p}| < k_F \\\\
0 \,, & \quad  |\textbf{p}| > k_F
\end{array}\,,\right.
\end{eqnarray}
with the normalization factor $\eta^{\textrm{pure}} = 1.389$. Here the proton momentum is constrained to be purely below the Fermi momentum. The red line in Fig.\,\ref{boundsigma3} represents the $\sigma_{\textrm{bound}}(|\textbf{k}|)$ obtained by this pure Fermi motion distribution in Eq.\,(\ref{modelbySargsianpure}).

As seen, the cross section obtained by SRC dominance model (blue line) reveals significant differences from both the free cross section (black line) and the cross section only considering Fermi motion (red line). This indicates that SRCs can dramatically alter the shape of the cross section, which provides a clear signature for this core structure in nucleus. As shown in Fig.\,\ref{boundsigma3}, the difference between cross sections is primarily caused by high-momentum protons in nuclei, which originate essentially from short-range hard interactions between nucleons---a fact has been confirmed in studies of nuclei ranging from beryllium to lead.

\textit{Feasibility at BESIII and STCF.}---In our study of quasielastic $\pi^+$-bound proton scattering, the proton targets are provided by the beryllium beam pipe of the BESIII detector, while the incident $\pi^+$ particles are sourced from charmonium decays. The BEIII has collected the world's largest samples of $J/\psi$ and $\psi(3686)$ events, comprising 10 billion and 2.7 billion events, respectively. Among the hundreds of known decay channels, numerous high-branching-fraction ($>10^{-3}$) channels decaying into pure charged final states, including a $\pi^+$, are ideal for this purpose. These other remaining charged tracks, originating directly from the interaction point, can be selected and identified with high efficiency.

Given the known four-momenta of the initial $J/\psi(\psi(3686))$ and the ``tagged" charged tracks, the $\pi^+$ can be reconstructed by requiring the recoiling mass of ``tagged'' charged tracks to agree with that of the pion. The momentum and direction of the pion can be determined as well, with uncertainties of a few MeV and a few milliradians, respectively. The resulting control samples are exceptionally pure (with a purity up to 99\%), as demonstrated in previous BESIII tracking efficiency studies. A simple estimate demonstrates the statistical capability of BESIII. Using published BESIII results for decays like $J/\psi \to K^+K^-\pi^+\pi^-$ and $J/\psi \to \pi^+\pi^-\pi^+\pi^-$ (with summed branching fractions of approximately $1.0 \times 10^{-2}$)~\cite{ParticleDataGroup:2024cfk} and tagging efficiency of $\sim$40\%, one can obtain the order of $4\times 10^{7}$ $\pi^+$ in the 10 billion $J/\psi$ event sample.
With these pions, and considering the typical pion-nucleon scattering cross section ($\sim$10\,mb) along with previous research on the material content in the BESIII beam pipe \cite{BESIII:2023clq}, we can expect to obtain $\sim10^4$ events of $\pi^+$-bound proton scattering.

The future Super $\tau$-Charm Facility (STCF) is designed to reach a peak luminosity of $0.5\times 10^{35} \, \rm{cm}^{-2}\rm{s}^{-1}$, which is 50 times higher than that of BEPCII.  Simply rescaling from the BESIII estimates demonstrates that the STCF would offer tremendous opportunities for the physics program discussed in this Letter.

\textit{Summary.}---SRCs are considered the core of nuclear structure, marking the transition from nucleonic to quark-gluon degrees of freedom. In this work, we identify and explore a promising avenue to probe this structure through the $\pi^+$-bound proton scattering at an electron-positron collider. We show that BESIII and the proposed STCF are feasible for this innovative study in nuclear and particle physics.

The study of atomic nuclei is driven by the ambition to map out the entire landscape from the nuclear radius ($\approx$5\,fm for a heavy nucleus) down to the scale of individual nucleons ($\approx$0.84\,fm). The internucleon distance marks a pivotal transition in this mapping. As a kind of important description of this transition, SRCs are thought to be keys to combine the particle physics perspective, where nuclei are seen as made up of quarks and gluons, with the traditional nuclear physics view that treats nuclei as collections of interacting nucleons. The proposal in this work can help to elucidate the structure of SRCs and enhance our understanding of fundamental aspects of QCD.

\textit{Acknowledgements.}---We thank Profs. Guang-Shun Huang, Shuang-Shi Fang, and Wen-Biao Yan for the inspiring discussions and suggestions on the measurements at BESIII. This work is supported in part by National Natural Science Foundation of China under Grant No. 12125503, 12335003, 12305106, 12475098, 12205255, and 12105247.

\begin{widetext}
\appendix
\section{Appendix A: The Fermi momentum of beryllium and the distribution of $E_\textrm{miss}$}
\label{App1}
The Fermi momentum is a cornerstone concept in nuclear physics which is directly related to the density of nuclear matter. For medium and heavy nuclei, the typical Fermi momentum approximately equal to 250\,MeV. In \cite{Tang:2002ww,Moniz:1971mt}, the authors have calculated $k_F=220\,\textrm{MeV}$ for $^{12}$C. In their work, the experimental data has been interpreted in terms of Fermi gas model, yielding the nuclear Fermi momentum as a function of mass number. The corresponding results are collected in Tab.\,\ref{kFvsA}.

\begin{table}[!htbp]
\renewcommand{\arraystretch}{1.5}
   \caption{Nuclear Fermi momentum $k_F \textrm{(MeV)}$ for different nuclei presented in \cite{Moniz:1971mt}.}\label{kFvsA}
	\begin{tabular}{c|c c c c c c c c c}
		\hline\hline
		\qquad~~~Nucleus \qquad~~~                 & ~~$^{6}$Li~~ & ~~$^{12}$C~~ & ~~$^{24}$Mg~~ & ~~$^{40}$Ca~~ & ~~$^{58.7}$Ni~~ & ~~$^{89}$Y~~ & ~~$^{118.7}$Sn~~ & ~~$^{181}$Ta~~ & ~~$^{208}$Pb~~                 \\\hline
		\qquad~~~$ k_F \textrm{(MeV)} $\qquad~~~   &  ~~$169$~~   & ~~$221$~~    & ~~$235$~~     & ~~$251$~~     & ~~$260$~~       & ~~$254$~~    & ~~$260$~~
                 & ~~$265$~~      & ~~$265$~~            \\
		\hline\hline
	\end{tabular}
\end{table}	

These values can be well reproduced by a polynomial fit, the obtained relationship between the mass number $A$ and the Fermi momentum $k_F$ is
\begin{eqnarray}
  k_F^A = -0.0039 A^2 + 1.10 A + 196.83 \,.
\end{eqnarray}
Therefore, it is reasonable to assign the Fermi momentum of beryllium to $k_F=206\,\textrm{MeV}$.

In beryllium, protons with momenta below $k_F=206\,\textrm{MeV}$ are well described by the mean field calculation, and their missing energy $E_{\textrm{miss}}\simeq 0$. Protons with momenta above $k_F$ are primarily contributed by SRCs. The description of their $E_{\textrm{miss}}$ can be inferred from experimental measurements on $^{12}$C \cite{CLAS:2020mom}, which revealed that when the proton momentum is small, the corresponding $E_{\textrm{miss}}$ is essentially zero. As the proton momentum increases, $E_{\textrm{miss}}$ begins to rise, which is dominated by the contributions from SRCs. The SRCs are structures with high locality and are insensitive to the properties of the other surrounding nucleons. Therefore these structures can be considered as ``universal'' (same for all kinds of nuclei). By fitting the experimental data, we find the distribution of the proton missing energy can be expressed in the following simple form:
\begin{eqnarray}\label{Emissfunction}
  E_{\textrm{miss}}(|\textbf{p}|)=\left\{\begin{array}{cc}
0 \,, & \qquad  |\textbf{p}|<k_F \\ \\
0.446 \, |\textbf{p}| -0.098  \,, & \qquad  |\textbf{p}|>k_F
\end{array}\,.\right.
\end{eqnarray}
This is useful for the calculation of cross section for $\pi^+ p$ scattering where the proton is bound in beryllium.

\section{Appendix B: Determination on $\sigma_{\textrm{free}}(\sqrt{s})$}
\label{App2}
The period from the 1960s to the present has seen the accumulation of extensive data for free $\pi^+ p$ scattering \cite{ParticleDataGroup:2024cfk}, as shown in Fig.\,\ref{Fitting_PDG_pionN}. Here we employ a data-fitting approach, three Breit-Wigner distributions have been used:

\begin{figure}[htbp]
\includegraphics[width=0.60\columnwidth]{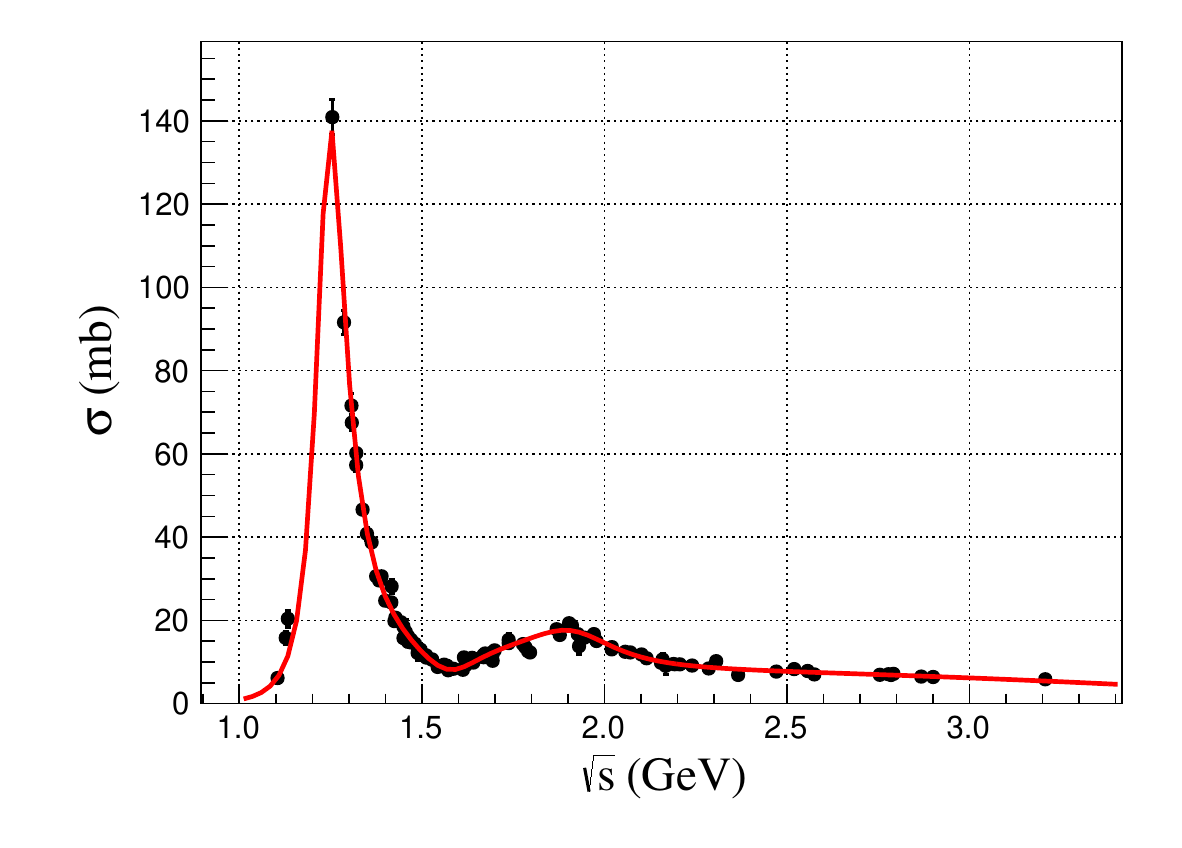}
\caption{The data on cross section for $\pi^+ p$ scattering as a function of $\sqrt{s}$, with the fitting result represented by the red solid line.}
\label{Fitting_PDG_pionN}
\end{figure}

\begin{eqnarray}
  \mathcal{A}_{BW(1232)} &=& \frac{a}{s-m^2(1232)+im(1232)\sigma(1232)} \,,\nn\\
  \mathcal{A}_{BW(1550)} &=& \frac{b}{s-m^2(1550)+im(1550)\sigma(1550)} \,,\nn\\
  \mathcal{A}_{BW(1910)} &=& \frac{c}{s-m^2(1910)+im(1910)\sigma(1910)} \,.
\end{eqnarray}
Therefore, the cross section is
\begin{eqnarray}
  \sigma_{\textrm{free}}(\sqrt{s}) = | e^{i \phi(1232)} \mathcal{A}_{BW(1232)} + e^{i \phi(1550)} \mathcal{A}_{BW(1550)} + e^{i \phi(1910)} \mathcal{A}_{BW(1910)} + d \sqrt{s} + f s |^2 \,.
\end{eqnarray}
The parameters are taken as:
\begin{eqnarray}
  && m(1232) = 1.235\,\textrm{GeV} \,, \quad \sigma(1232) = 0.091\,\textrm{GeV} \,, \quad \phi(1232) = -3.275\,, \quad a = -1.415\,\textrm{GeV} \,, \nn\\
  && m(1550) = 1.569\,\textrm{GeV} \,, \quad \sigma(1550) = 0.200\,\textrm{GeV} \,, \quad \phi(1550) = -1.778\,, \quad b = 0.385\,\textrm{GeV} \,, \nn\\
  && m(1910) = 1.911\,\textrm{GeV} \,, \quad \sigma(1910) = 0.302\,\textrm{GeV} \,, \quad \phi(1910) = -1.514\,, \quad c = -0.718\,\textrm{GeV} \,, \nn\\
  && d = 2.161\,\textrm{GeV}^{-2} \,, \quad f = -0.466\,\textrm{GeV}^{-3} \,.
\end{eqnarray}
The result is represented by the red solid line in Fig.\,\ref{Fitting_PDG_pionN}, which will be substituted into Eq.\,(\ref{sigmabound1}) to calculate the $\sigma_{\textrm{bound}}(|\textbf{k}|)$.

\section{Appendix C: Momentum distributions of protons in beryllium}
\label{App3}
In this work, we utilize the SRC dominance model to describe the momentum distributions of protons bound in beryllium. According to this model, short-range component of the nucleon-nucleon potential is responsible for hard interactions between nucleons, which pushes them from low momenta to high momenta (well in excess of the Fermi momentum). Therefore, the momentum distribution of proton in nuclei should be divided into two parts based on its magnitude,
\begin{eqnarray}\label{modelbySargsianAPP}
  \rho_p(|\textbf{p}|)=\left\{\begin{array}{cc}
\eta \cdot \rho_p^{\textrm{MF}}(|\textbf{p}|) \,, & \quad  |\textbf{p}| < k_F \\\\
\frac{A}{2 Z} \cdot a_2(\textrm{Be}/d) \cdot \rho_d(|\textbf{p}|) \,, & \quad  |\textbf{p}| > k_F
\end{array}\,,\right.
\end{eqnarray}
here $A=9$ ($Z=4$) is the mass (proton) number of beryllium and $k_F=0.206\,\textrm{GeV}$ is the Fermi momentum. We will now proceed to analyze the remaining quantities in Eq.\,(\ref{modelbySargsianAPP}) one by one.
\begin{itemize}
  \item The mean field momentum distribution $\rho_p^{\textrm{MF}}(|\textbf{p}|)$ for different nuclei has been calculated with the Woods-Saxon (WS) wave functions. A Boltzmann distribution was presented to describe the distribution \cite{Vanhalst:2012ur}
\begin{eqnarray}\label{Boltzmanndistribution}
  \rho_p^{\textrm{MF}}(|\textbf{p}|) = \left(\frac{2\pi}{m_N p_T}\right)^{3/2} \, e^{-\frac{|\textbf{p}|^2}{2m_N p_T}} \,,
\end{eqnarray}
in which the Boltzmann parameters were determined as $p_T \approx 12\,\textrm{MeV}$ for $^{12}$C, $p_T \approx 14\,\textrm{MeV}$ for $^{56}$Fe, and $p_T \approx 16\,\textrm{MeV}$ for $^{208}$Pb. These results allow us to evaluate the Boltzmann parameter for $^{9}$Be to be $p_T \approx 11.8\,\textrm{MeV}$. Fig.\,\ref{BoltzmannFit} shows the mean field momentum distribution $\rho_p^{\textrm{MF}}(|\textbf{p}|)$ for $^{9}$Be.

\begin{figure}[htbp]
\includegraphics[width=0.55\columnwidth]{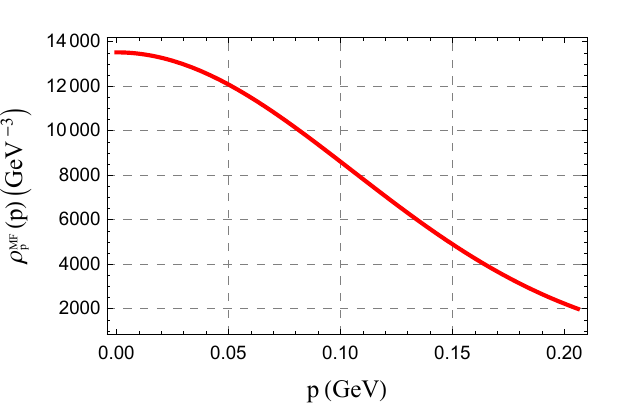}
\caption{The mean field proton momentum distribution $\rho_p^{\textrm{MF}}(|\textbf{p}|)$ for $^{9}$Be obtained from the Boltzmann distribution in Eq.\,(\ref{Boltzmanndistribution}).}
\label{BoltzmannFit}
\end{figure}

  \item The SRC scale factor $a_2(\textrm{Be}/d)$ characterizes the probability of finding a high-momentum nucleon in beryllium relative to deuteron. Here we take the measurement $a_2(\textrm{Be}/d) = 4.40$ in \cite{CLAS:2019vsb}. Due to the enhancement of this factor, the proton distribution does not drop rapidly at high momentum; instead, it develops a broad shoulder (ridge).

  \item The proton momentum distribution in the deuteron $\rho_d(|\textbf{p}|)$ has been calculated through a Hamiltonian containing the Argonne v$_{18}$ two-nucleon potential in \cite{Wiringa:2013ala}. Its result is shown in Fig.\,\ref{FittingAV18UXDeu}, as previously introduced, here we are concerned with values where $|\textbf{p}| > 0.206\,\textrm{GeV}$ (i.e. $|\textbf{p}| > 1.044\,\textrm{fm}^{-1}$).

\begin{figure}[htbp]
\includegraphics[width=0.55\columnwidth]{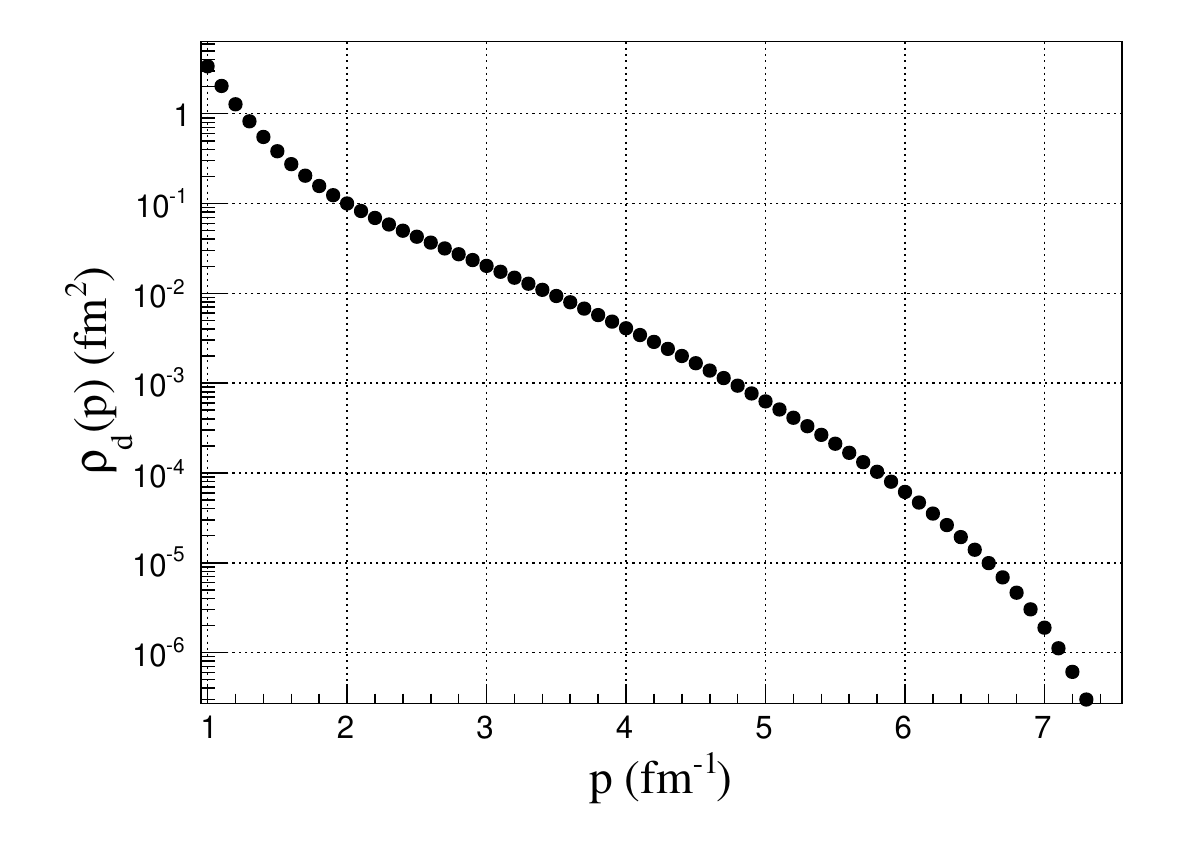}
\caption{The proton momentum distribution in deuteron with $|\textbf{p}| > 1.044\,\textrm{fm}^{-1}$.}
\label{FittingAV18UXDeu}
\end{figure}

  \item We substitute all the above results into Eq.\,(\ref{modelbySargsianAPP}) and then determine the value of $\eta$ through the normalization condition
     \begin{eqnarray}
       \frac{4\pi}{(2\pi)^3}\int_0^\infty \rho_{p}(|\textbf{p}|) |\textbf{p}|^2 d|\textbf{p}| =1 \,.
     \end{eqnarray}
     This gives us $\eta=0.8796$.

\end{itemize}

Thus, we have arrived at the complete expression for the proton momentum distribution $\rho_p(|\textbf{p}|)$ in beryllium.

\end{widetext}


\begin{thebibliography}{}

\bibitem{Frankfurt:1988nt}
L.~L.~Frankfurt and M.~I.~Strikman,
Phys. Rept. \textbf{160}, 235-427 (1988)
doi:10.1016/0370-1573(88)90179-2

\bibitem{Arrington:2011xs}
J.~Arrington, D.~W.~Higinbotham, G.~Rosner and M.~Sargsian,
Prog. Part. Nucl. Phys. \textbf{67}, 898-938 (2012)
doi:10.1016/j.ppnp.2012.04.002
[arXiv:1104.1196 [nucl-ex]].

\bibitem{Hen:2016kwk}
O.~Hen, G.~A.~Miller, E.~Piasetzky and L.~B.~Weinstein,
Rev. Mod. Phys. \textbf{89}, no.4, 045002 (2017)
doi:10.1103/RevModPhys.89.045002
[arXiv:1611.09748 [nucl-ex]].

\bibitem{Miller:2020eyc}
G.~A.~Miller,
Phys. Rev. C \textbf{102}, no.5, 055206 (2020)
doi:10.1103/PhysRevC.102.055206
[arXiv:2008.06524 [nucl-th]].

\bibitem{Frankfurt:1993sp}
L.~L.~Frankfurt, M.~I.~Strikman, D.~B.~Day and M.~Sargsian,
Phys. Rev. C \textbf{48}, 2451-2461 (1993)
doi:10.1103/PhysRevC.48.2451

\bibitem{Tang:2002ww}
A.~Tang, J.~W.~Watson, J.~L.~S.~Aclander, J.~Alster, G.~Asryan, Y.~Averichev, D.~Barton, V.~Baturin, N.~Bukhtoyarova and A.~Carroll, \textit{et al.}
Phys. Rev. Lett. \textbf{90}, 042301 (2003)
doi:10.1103/PhysRevLett.90.042301
[arXiv:nucl-ex/0206003 [nucl-ex]].

\bibitem{Hen:2014nza}
O.~Hen, M.~Sargsian, L.~B.~Weinstein, E.~Piasetzky, H.~Hakobyan, D.~W.~Higinbotham, M.~Braverman, W.~K.~Brooks, S.~Gilad and K.~P.~Adhikari, \textit{et al.}
Science \textbf{346}, 614-617 (2014)
doi:10.1126/science.1256785
[arXiv:1412.0138 [nucl-ex]].

\bibitem{Subedi:2008zz}
R.~Subedi, R.~Shneor, P.~Monaghan, B.~D.~Anderson, K.~Aniol, J.~Annand, J.~Arrington, H.~Benaoum, W.~Bertozzi and F.~Benmokhtar, \textit{et al.}
Science \textbf{320}, 1476-1478 (2008)
doi:10.1126/science.1156675
[arXiv:0908.1514 [nucl-ex]].

\bibitem{Fomin:2011ng}
N.~Fomin, J.~Arrington, R.~Asaturyan, F.~Benmokhtar, W.~Boeglin, P.~Bosted, A.~Bruell, M.~H.~S.~Bukhari, M.~E.~Christy and E.~Chudakov, \textit{et al.}
Phys. Rev. Lett. \textbf{108}, 092502 (2012)
doi:10.1103/PhysRevLett.108.092502
[arXiv:1107.3583 [nucl-ex]].

\bibitem{CLAS:2020mom}
A.~Schmidt \textit{et al.} [CLAS],
Nature \textbf{578}, no.7796, 540-544 (2020)
doi:10.1038/s41586-020-2021-6
[arXiv:2004.11221 [nucl-ex]].

\bibitem{CLAS:2005ola}
K.~S.~Egiyan \textit{et al.} [CLAS],
Phys. Rev. Lett. \textbf{96}, 082501 (2006)
doi:10.1103/PhysRevLett.96.082501
[arXiv:nucl-ex/0508026 [nucl-ex]].

\bibitem{Li:2022fhh}
S.~Li, R.~Cruz-Torres, N.~Santiesteban, Z.~H.~Ye, D.~Abrams, S.~Alsalmi, D.~Androic, K.~Aniol, J.~Arrington and T.~Averett, \textit{et al.}
Nature \textbf{609}, no.7925, 41-45 (2022)
doi:10.1038/s41586-022-05007-2
[arXiv:2210.04189 [nucl-ex]].

\bibitem{Zhang:2025nst}
Y.~P.~Zhang, Z.~H.~Ye, D.~Nguyen, P.~Aguilera, Z.~Ahmed, H.~Albataineh, K.~Allada, B.~Anderson, D.~Anez and K.~Aniol, \textit{et al.}
[arXiv:2504.17462 [nucl-ex]].

\bibitem{Xu:2019wso}
J.~Xu and F.~Yuan,
Phys. Lett. B \textbf{801}, 135187 (2020)
doi:10.1016/j.physletb.2019.135187
[arXiv:1908.10413 [hep-ph]].

\bibitem{Hatta:2019ocp}
Y.~Hatta, M.~Strikman, J.~Xu and F.~Yuan,
Phys. Lett. B \textbf{803}, 135321 (2020)
doi:10.1016/j.physletb.2020.135321
[arXiv:1911.11706 [hep-ph]].

\bibitem{Ma:2023tsi}
N.~N.~Ma, T.~F.~Wang and R.~Wang,
Phys. Rev. C \textbf{108}, no.6, 6 (2023)
doi:10.1103/PhysRevC.108.065203
[arXiv:2305.18112 [nucl-th]].

\bibitem{Sargsian:2012sm}
M.~M.~Sargsian,
Phys. Rev. C \textbf{89}, no.3, 034305 (2014)
doi:10.1103/PhysRevC.89.034305
[arXiv:1210.3280 [nucl-th]].

\bibitem{Wang:2024ikx}
W.~Wang, J.~Xu, X.~H.~Yang, Y.~T.~Zhang and S.~Zhao,
[arXiv:2409.14367 [hep-ph]].

\bibitem{Detmold:2020snb}
W.~Detmold \textit{et al.} [NPLQCD],
Phys. Rev. Lett. \textbf{126}, no.20, 202001 (2021)
doi:10.1103/PhysRevLett.126.202001
[arXiv:2009.05522 [hep-lat]].

\bibitem{Chen:2024rgi}
C.~Chen \textit{et al.} [CLQCD and Lattice Parton],
Phys. Rev. D \textbf{111}, no.7, 074506 (2025)
doi:10.1103/PhysRevD.111.074506
[arXiv:2408.12819 [hep-lat]].

\bibitem{EuropeanMuon:1983wih}
J.~J.~Aubert \textit{et al.} [European Muon],
Phys. Lett. B \textbf{123}, 275-278 (1983)
doi:10.1016/0370-2693(83)90437-9

\bibitem{EuropeanMuon:1988lbf}
J.~Ashman \textit{et al.} [European Muon],
Phys. Lett. B \textbf{202}, 603-610 (1988)
doi:10.1016/0370-2693(88)91872-2

\bibitem{Weinstein:2010rt}
L.~B.~Weinstein, E.~Piasetzky, D.~W.~Higinbotham, J.~Gomez, O.~Hen and R.~Shneor,
Phys. Rev. Lett. \textbf{106}, 052301 (2011)
doi:10.1103/PhysRevLett.106.052301
[arXiv:1009.5666 [hep-ph]].

\bibitem{CLAS:2019vsb}
B.~Schmookler \textit{et al.} [CLAS],
Nature \textbf{566}, no.7744, 354-358 (2019)
doi:10.1038/s41586-019-0925-9
[arXiv:2004.12065 [nucl-ex]].

\bibitem{CLAS:2018xvc}
M.~Duer \textit{et al.} [CLAS],
Phys. Rev. Lett. \textbf{122}, no.17, 172502 (2019)
doi:10.1103/PhysRevLett.122.172502
[arXiv:1810.05343 [nucl-ex]].

\bibitem{Chen:2016bde}
J.~W.~Chen, W.~Detmold, J.~E.~Lynn and A.~Schwenk,
Phys. Rev. Lett. \textbf{119}, no.26, 262502 (2017)
doi:10.1103/PhysRevLett.119.262502
[arXiv:1607.03065 [hep-ph]].

\bibitem{Wang:2020uhj}
X.~G.~Wang, A.~W.~Thomas and W.~Melnitchouk,
Phys. Rev. Lett. \textbf{125}, 262002 (2020)
doi:10.1103/PhysRevLett.125.262002
[arXiv:2004.03789 [hep-ph]].

\bibitem{Arrington:2012ax}
J.~Arrington, A.~Daniel, D.~Day, N.~Fomin, D.~Gaskell and P.~Solvignon,
Phys. Rev. C \textbf{86}, 065204 (2012)
doi:10.1103/PhysRevC.86.065204
[arXiv:1206.6343 [nucl-ex]].

\bibitem{Malace:2014uea}
S.~Malace, D.~Gaskell, D.~W.~Higinbotham and I.~Cloet,
Int. J. Mod. Phys. E \textbf{23}, no.08, 1430013 (2014)
doi:10.1142/S0218301314300136
[arXiv:1405.1270 [nucl-ex]].

\bibitem{Wang:2024cpx}
W.~Wang, J.~Xu, X.~H.~Yang and S.~Zhao,
Eur. Phys. J. A \textbf{61}, no.5, 112 (2025)
doi:10.1140/epja/s10050-025-01588-4
[arXiv:2401.16662 [hep-ph]].

\bibitem{nCTEQ:2023cpo}
A.~W.~Denniston \textit{et al.} [nCTEQ],
Phys. Rev. Lett. \textbf{133}, no.15, 152502 (2024)
doi:10.1103/PhysRevLett.133.152502
[arXiv:2312.16293 [hep-ph]].

\bibitem{Bernard:1995dp}
V.~Bernard, N.~Kaiser and U.~G.~Meissner,
Int. J. Mod. Phys. E \textbf{4}, 193-346 (1995)
doi:10.1142/S0218301395000092
[arXiv:hep-ph/9501384 [hep-ph]].

\bibitem{Ellis:1997kc}
P.~J.~Ellis and H.~B.~Tang,
Phys. Rev. C \textbf{57}, 3356-3375 (1998)
doi:10.1103/PhysRevC.57.3356
[arXiv:hep-ph/9709354 [hep-ph]].

\bibitem{Bernard:2007zu}
V.~Bernard,
Prog. Part. Nucl. Phys. \textbf{60}, 82-160 (2008)
doi:10.1016/j.ppnp.2007.07.001
[arXiv:0706.0312 [hep-ph]].

\bibitem{BESIII:2009fln}
M.~Ablikim \textit{et al.} [BESIII],
Nucl. Instrum. Meth. A \textbf{614}, 345-399 (2010)
doi:10.1016/j.nima.2009.12.050
[arXiv:0911.4960 [physics.ins-det]].

\bibitem{Yuan:2021yks}
C.~Z.~Yuan and M.~Karliner,
Phys. Rev. Lett. \textbf{127}, no.1, 012003 (2021)
doi:10.1103/PhysRevLett.127.012003
[arXiv:2103.06658 [hep-ex]].

\bibitem{BESIII:2023clq}
M.~Ablikim \textit{et al.} [BESIII],
Phys. Rev. Lett. \textbf{130}, no.25, 251902 (2023)
doi:10.1103/PhysRevLett.130.251902
[arXiv:2304.13921 [hep-ex]].

\bibitem{BESIII:2023trh}
M.~Ablikim \textit{et al.} [BESIII],
Phys. Rev. C \textbf{109}, no.5, L052201 (2024)
doi:10.1103/PhysRevC.109.L052201
[arXiv:2310.00720 [nucl-ex]].

\bibitem{BESIII:2024geh}
M.~Ablikim \textit{et al.} [BESIII],
Phys. Rev. Lett. \textbf{132}, no.23, 231902 (2024)
doi:10.1103/PhysRevLett.132.231902
[arXiv:2401.09012 [hep-ex]].

\bibitem{CiofidegliAtti:1995qe}
C.~Ciofi degli Atti and S.~Simula,
Phys. Rev. C \textbf{53}, 1689 (1996)
doi:10.1103/PhysRevC.53.1689
[arXiv:nucl-th/9507024 [nucl-th]].

\bibitem{CiofidegliAtti:2007ork}
C.~Ciofi degli Atti, L.~L.~Frankfurt, L.~P.~Kaptari and M.~I.~Strikman,
Phys. Rev. C \textbf{76}, 055206 (2007)
doi:10.1103/PhysRevC.76.055206
[arXiv:0706.2937 [nucl-th]].

\bibitem{Kim:2024wne}
D.~N.~Kim, O.~Hen, G.~A.~Miller, E.~Piasetzky, M.~Strikman and L.~Weinstein,
Phys. Rev. C \textbf{111}, no.6, 065201 (2025)
doi:10.1103/xfkg-3m6q
[arXiv:2404.15442 [nucl-th]].

\bibitem{Xing:2023uhj}
W.~Xing, X.~G.~Wang and A.~W.~Thomas,
Phys. Lett. B \textbf{846}, 138195 (2023)
doi:10.1016/j.physletb.2023.138195
[arXiv:2305.13666 [hep-ph]].

\bibitem{ParticleDataGroup:2024cfk}
S.~Navas \textit{et al.} [Particle Data Group],
Phys. Rev. D \textbf{110}, no.3, 030001 (2024)
doi:10.1103/PhysRevD.110.030001


\bibitem{Moniz:1971mt}
E.~J.~Moniz, I.~Sick, R.~R.~Whitney, J.~R.~Ficenec, R.~D.~Kephart and W.~P.~Trower,
Phys. Rev. Lett. \textbf{26}, 445-448 (1971)
doi:10.1103/PhysRevLett.26.445

\bibitem{Vanhalst:2012ur}
M.~Vanhalst, J.~Ryckebusch and W.~Cosyn,
Phys. Rev. C \textbf{86}, 044619 (2012)
doi:10.1103/PhysRevC.86.044619
[arXiv:1206.5151 [nucl-th]].

\bibitem{Wiringa:2013ala}
R.~B.~Wiringa, R.~Schiavilla, S.~C.~Pieper and J.~Carlson,
Phys. Rev. C \textbf{89}, no.2, 024305 (2014)
doi:10.1103/PhysRevC.89.024305
[arXiv:1309.3794 [nucl-th]].



\end{thebibliography}
\end{document}